\documentclass{aa}
\usepackage{graphicx}
\usepackage{amsmath,amssymb}
\usepackage{txfonts}
\usepackage{natbib}
\begin{document}
   \title{Dust-driven winds and mass loss of C-rich AGB stars with
          subsolar metallicities}

%   \subtitle{I. Subtitle}

   \author{
           A.~Wachter \inst{1,2}
       \and
           J.~M.~Winters \inst{3}
       \and
           K.-P.~Schr{\"o}der \inst{4}
       \and
           E.~Sedlmayr \inst{1}
          }

   \offprints{A.~Wachter}

   \institute{
              Zentrum f{\"u}r Astronomie und Astrophysik (ZAA),
              Technische Universit{\"a}t Berlin, Hardenbergstr.~36,
              10623 Berlin, Germany
              \\ \email{awachter@physik.tu-berlin.de}
         \and
             Department of Physics and Astronomy, Division of Astronomy and
             Space Physics, Uppsala University, Box 515, 75120 Uppsala, Sweden
         \and
             Institut de RadioAstronomie Millim{\'e}trique (IRAM), 300
             rue de la Piscine, Domaine Universitaire, 38406 Saint
             Martin d'H{\`e}res, France
%             \\ \email{winters@iram.fr}
         \and
             Departamento de Astronomia de la Universidad de
             Guanajuato, Apartado Postal 144, C.P. 36000, Guanajuato, 
             GTO, Mexico
%            \\ \email{kps@astro.ugto.mx}
             }

   \date{Received 02 April 2008 / Accepted 21 May 2008}

% \abstract{}{}{}{}{} 
% 5 {} token are mandatory
 
  \abstract
   {}
   {We investigate the mass loss of highly evolved, low- and
   intermediate mass stars and stellar samples with subsolar
   metallicity. We give a qualitative as well as quantitative
   description which can be applied to LMC/SMC-type stellar
   populations.}
   {For that purpose we apply the same approach as we did for solar
   metallicity stars and calculate hydrodynamical wind models including
   dust formation with LMC and SMC abundances under consideration of an
   adapted model assumption. In particular, we improved the treatment
   of the radiative transfer problem in order to accommodate larger
   non-local contributions occurring with smaller opacities. For each
   wind model we determine an averaged mass-loss rate. The resulting,
   approximate mass-loss formulae are then applied to well-tested and
   calibrated stellar evolution calculations in order to quantify the
   stellar mass loss.}
   {The dynamical models for LMC and SMC metallicity result in mass-loss
   rates of the same order of magnitude as the solar metallicity models
   which is in this basic approach in agreement with observations. The
   hydrodynamical properties like e.g.\ the outflow velocity differ (for
   fixed C/O abundance ratio) noticeably, though. While critical
   luminosities of LMC and solar metallicity models fairly coincide, the
   SMC models need higher luminosities to develop dust-driven winds.
   }
   {}

   \keywords{Hydrodynamics --
%             Radiative transfer  --
             Stars: AGB and post-AGB --
%             Stars: atmospheres --
             Stars: carbon --
%             circumstellar matter --
             Stars: evolution --
%             Hertzsprung-Russell (HR) and C-M diagrams --
             Stars: late-type --
             Stars: mass-loss
            }

   \maketitle
%
%_______________________________________________________________________

\section{Introduction}
On the asymptotic giant branch (AGB) stars with low- and intermediate
main sequence mass ($\sim$1 $\dots$ 8~$M_\odot$, precise values depend
on the initial element abundances / metallicity and are as well model
dependent) are in their late stage of evolution. This phase is
characterised by instabilities leading to stellar pulsations as observed
in long-period variables (LPVs) or Mira stars. \citet{cmlah2001}, for
instance, found that 65\% of the stars in the considered sample of
giants in the Large Magellanic Cloud (LMC) are LPVs. Furthermore, this
phase is connected to strong mass-loss of more than a few $10^{-5}\,
M_\odot$~yr$^{-1}$. By means of their mass loss, these stars contribute
significantly to the enrichment of the interstellar matter with material
processed through nuclear burning reactions in the form of molecules and
dust, and are therefore not only interesting as objects of stellar
evolution but also in terms of galactic chemical evolution. In this
paper we extend our set of models with solar element composition (apart
from the C/O ratio) to subsolar metallicities. We are using abundances
as observed in the Magellanic Clouds \citep[see table~A.1
in][]{hws2000}. The metallicity of the Large Magellanic Cloud (LMC) is
$Z=0.008$, i.e.\ roughly a good third of the solar value, while the
Small Magellanic Cloud (SMC) has on average an even lower metallicity of
$Z=0.004$.

%__________________________________________________________________

\section{Dust-driven Wind Models}

We base our description of the mass loss of late AGB stars with less
than solar metallicities on time dependent hydrodynamical wind models
that include the formation of dust grains which, by radiation pressure,
drive the massive outflows. These models have been developed in the last
decade by \citet{ggs1990}, \citet{fgs1992}, \citet{wfgs1994,wljhs2000},
\citet{jwls2003}.

For (apart from the C/O abundance ratio) solar abundances, these models
yield time averaged mass loss rates of up to a few $10^{-5}$ M$_\odot$
yr$^{-1}$, both for oxygen-rich and carbon-rich mixtures. This is
consistent with rates inferred from, e.g.\ CO rotational line
observations of AGB stars in the solar neighbourhood.

The models are obtained from the solution of the non-linearly coupled
system of equations describing the hydrodynamical and thermodynamical
structure of a spherically symmetric, pulsating circumstellar shell, its
chemical composition, and the nucleation, growth, and evaporation of
dust grains. The hydrodynamical structure (mass density $\rho$ and
outflow velocity $v$) follows from the equation of continuity and the
equation of motion which includes radiation pressure on dust grains. The
law of energy conservation and radiative transfer determine the
temperature structure. In this work we consider models with carbon-rich
composition assuming that oxygen is locked in the CO molecule and
assuming chemical equilibrium. The formation, growth, and evaporation of
carbon grains is calculated according to the moment method developed by
\citet{gs1988} and \citet{ggs1990}.

The equations are discretised in Lagrangian coordinates, i.e.\ in a
co-moving coordinate system. A detailed description of the
transformation and discretisation is given in \citet{fgs1992} and
\citet{fleis1994}.
Furthermore, a summary of the latest physical assumptions can be found
in \citet{sws2003}.

When applied to winds with subsolar metallicities these models show a
rather quasi-static nature \citep{has2002}.
As already pointed out by these authors, this is not consistent with
observations of AGB stars, e.g.\ in the Magellanic Clouds which indicate
mass-loss rates of the same order of magnitude as they are observed from
stars with solar element abundances. Therefore, we reconsidered the
model assumptions. In particular, we improved the treatment of the
radiative transfer as follows:

In the solar metallicity case the shell has been considered to be
optically thick so that the flux-averaged dust extinction is determined
by the Planck-mean absorption coefficient at the local equilibrium
temperature. However, if the atmosphere is considered to be partly
transparent, the dust grains are to some extent exposed to the direct
radiation from the stellar photosphere. Therefore, the flux-mean dust
opacity for lower metallicity is represented by assuming a {\it
non}-local radiation field, which is characterised by a temperature $T$
determined according to
%\begin{equation}
$
  T^4 = T_\star^4 e^{-\tau} + T_{\text{eq}}^4 (1 - e^{-\tau})
$,
%\end{equation}
i.e.\ the radiation field is interpolated in optical depth $\tau$
between the local equilibrium radiation and the stellar (black-body)
radiation field (represented by $T_\text{eq}$ and $T_\star$,
respectively).

The gas opacity $\kappa_\text{g}$/$\rho$ is set to a constant value of
$2 \times 10^{-4}$~cm$^{2}$~g$^{-1}$ in the solar models. From figure~2
of \citet{hws2000} it is clear that the mean gas opacity drops with
lower metallicity. Therefore, we reduced the value of the gas opacity
in the LMC and SMC models to $1 \times 10^{-4}$, and $0.5 \times
10^{-4}$~cm$^2$~g$^{-1}$, respectively.

%______________________________________________________________

\section{Results}

Since the main purpose of this work is to derive mass-loss rates for
subsolar metallicities, the focus when choosing the model parameters was
primarily on covering the range of stellar parameters $M$, $L$, and $T$
provided by the stellar evolution code for each fixed element
composition (LMC/SMC) than on performing a detailed parameter
study. Nevertheless, the dependence on the carbon overabundance is
shortly discussed in the following. Table~\ref{tab:parRanges} summarises
the parameter ranges of our subsolar metallicity model sets.

\begin{table}[tbp]
\caption{Essential input parameter ranges covered by our collection of subsolar
  metallicity models.}
\label{tab:parRanges}
\centering
\begin{tabular}{l||l|l}
type                                  & LMC                & SMC                \\ \hline
$L_\star$ [$L_\odot$]                 & $5000 \dots 15000$ & $7000 \dots 12000$ \\
$T_\star$ [K]                         & $2200 \dots 3200$  & $2400 \dots 3000$  \\
$M_\star$ [$M_\odot$]                 & $0.7 \dots 1.0$    & $0.6 \dots 1.1$    \\
$\epsilon_{\rm C} / \epsilon_{\rm O}$ & $1.3 / 1.8$        & $1.8$              \\
$P$ [d]                               & $325  \dots 650$   & $450  \dots 650$   \\
$\Delta v$ [km/s]                     & $2 / 5$            & $5$                \\
\end{tabular}
\end{table}

Concerning the radial structure of our wind models there basically exist
two different types of models:

A) models where effective dust formation and growth takes place. In this
class of models the radiative acceleration on dust exceeds the local
gravitation in the region close to the stellar photosphere (typically
inside of $\approx 5 R_{\star}$). In this case, the matter is lifted out
of the stars' gravitational field by radiation pressure on dust, and a
layered structure of the hydrodynamical and the dust quantities results
in the inner part of the circumstellar shell (see, e.g.,
figure~\ref{fig:SMC-LMC-solar}). Generally this type of models shows
high mass-loss rates ($>$ a few $10^{-7}M_\odot$yr$^{-1}$) and outflow
velocities in excess of $\approx 5\,$km\,s$^{-1}$.

B) models showing a smooth, quasi-stationary wind structure. In these
model, radiation pressure on dust is not sufficient to lift the matter
out of the stars' gravitational field, the resulting mass loss rates
(and outflow velocities) are significantly smaller than in the type A
models. 

This separation has been discussed in detail in
\citet{wljhs2000} for the case of solar element abundances.

As the distinction between A-type and B-type models is due to the ratio
of radiative to gravitational acceleration (larger or smaller than
unity) in the inner shell region, the same type of separation is found
for sub-solar metallicities. For the SMC and LMC models, this separation
is later on considered in terms of a ``critical luminosity''.

In order to characterise the outflow of each time dependent wind model
time-averages of several quantities are calculated. The most obvious
quantities for this purpose are the mass-loss rate $\langle \dot M
\rangle$ and the final outflow velocity $\langle v_\infty
\rangle$. Furthermore we consider the dust-to-gas ratio $\langle
\rho_{\mathrm d}/\rho_{\mathrm g} \rangle$ and the ratio of radiative
to gravitational acceleration $\langle \alpha \rangle$, since these
are expected to be significant for the general behaviour of the wind.

To calculate the averages the same procedure as used in the solar
metallicity models is followed, i.e.\ we first average $\dot M$, $v$,
and $\rho_\text{d/g}$ in the radial coordinate between 40 and 60
$R_\star$ and then over time, typically 20 periods for LMC and 80
periods for SMC models. For $\langle \alpha \rangle$ a different
procedure is followed, though. It is a time-average only, namely the
average of the ratio of radiative to gravitational acceleration at the
radius of the first maximum of the condensation degree, i.e.\ in the
innermost dust layer (see, e.g., the innermost peak of $\alpha$ in the
second panel of figure~\ref{fig:SMC-LMC-solar}).

\subsection{Influence of the C/O ratio}\label{sec:COratio}

The C/O abundance ratio determines how much carbon is available for dust
formation, since we consider carbon rich chemistry and total
CO-blocking. As far as the derivation of a mass loss formula applicable
for stellar evolution is concerned, the solar metallicity models already
demonstrated that the mass loss rate does not depend heavily on this
model parameter \citep[see][]{afs1997,wswas2002}.
Instead it can be seen as a critical parameter in the sense that the
mass loss rate remains the same order of magnitude once the C/O ratio is
above a certain threshold.

The results of a set of calculations for LMC and SMC abundances with
varying C/O ratio are summarised in table~\ref{tab:COvariation}. In the
LMC models with $\epsilon_{\rm C}/\epsilon_{\rm O}$ = 1.3 and 1.5 the
material expands so slowly that after 110 and 90 periods the outermost
grid point has just reached 10 and 25 stellar radii, respectively. This
is well below the region of 40 to 60 stellar radii where we generally
take the averages. The calculations have not been followed any further
since the models up to that instant show a virtually stationary radial
structure.

\begin{table}[tbp]
  \caption{Variation of the carbon overabundance $\epsilon_\text{C/O}$
    for otherwise fixed parameters.}
  \label{tab:COvariation}
  \centering
\begin{tabular}{lr|crrrr}
 & $\epsilon_\text{C/O}$ & periods & $|\dot M|$ & $\langle \alpha \rangle$ & $v_\infty$ & $\rho_\text{d/g}$ \\ \hline \hline
 \multicolumn{7}{l}{}\\
 \multicolumn{2}{l|}{type: LMC} & \multicolumn{5}{l}{$M = 0.8\, M_\odot$, $T = 2800$~K, $L = 15000\, L_\odot$,} \\
 & & \multicolumn{5}{l}{$P = 400$~d, $\Delta v = 5$~km~s$^{-1}$}\\ \hline
 & 1.30 & \multicolumn{5}{l}{Expansion after 110~$P$: $<$ 10~R$_\star$} \\
 & 1.50 & \multicolumn{5}{l}{Expansion after $\phantom{1}$90~$P$: $<$ 25~R$_\star$} \\ \cline{1-7}
 & 1.70 &  70--90  & 3.51e-05 & 6.10 & 18.83 & 2.01e-3 \\
 & 1.80 &  70--90  & 3.30e-05 & 3.92 & 21.04 & 2.68e-3 \\
 & 2.50 &  70--90  & 4.58e-05 & 6.03 & 30.16 & 7.13e-3 \\ \hline
 \multicolumn{7}{l}{}\\ %\hline
 \multicolumn{2}{l|}{type: SMC} & \multicolumn{5}{l}{$M = 1.0\, M_\odot$, $T = 2600$~K, $L = 11000\, L_\odot$,} \\
 & & \multicolumn{5}{l}{$P = 650$~d, $\Delta v = 5$~km~s$^{-1}$}\\ \hline
 & 1.30 & \multicolumn{5}{l}{Expansion after 150~$P$: $<$ 35~R$_\star$} \\
$\alpha < 1$ & 1.50 & 70--150 & 7.17e-07 & 0.06 & 3.23  & 1.69e-4 \\
             & 1.70 & 70--150 & 4.74e-06 & 0.11 & 4.95  & 1.17e-3 \\  \cline{1-7}
             & 1.80 & 70--150 & 3.69e-05 & 1.26 & 8.56  & 1.82e-3 \\
\raisebox{1.5ex}[-1.5ex]{$\alpha > 1$} & 2.00 &  70--150 & 4.55e-05 & 2.19 & 10.61 & 2.75e-3 \\ \hline
 \multicolumn{7}{l}{}\\
 \multicolumn{2}{l|}{type: SMC} & \multicolumn{5}{l}{$M = 0.8\, M_\odot$, $T = 2800$~K, $L = 10000\, L_\odot$,} \\
 & & \multicolumn{5}{l}{$P = 600$~d, $\Delta v = 5$~km~s$^{-1}$} \\ \hline
 & 1.30 & 130--150 & 1.24e-07 & 0.27 & 1.91  & 1.93e-9 \\
\raisebox{1.5ex}[-1.5ex]{$\alpha < 1$} & 1.50 &  70--150 & 8.90e-07 & 0.21 & 3.75  & 4.74e-4 \\ \cline{1-7}
             & 1.70 &  70--150 & 1.72e-05 & 1.02 &  8.31 & 1.27e-3 \\
$\alpha > 1$ & 1.80 &  70--150 & 2.88e-05 & 2.04 & 10.57 & 1.75e-3 \\
             & 2.00 &  70--150 & 3.76e-05 & 3.11 & 12.10 & 2.89e-3 
\end{tabular}
\end{table}
%%%%%%%%%%%%%%%%%%%%%%%%%%%%%%%%%%%%%%%%%%%%%%%%%%%%%%%%%%%%%%%%%%%%%%%%

\subsection{Comparison of models with different element abundances}
\label{sec:sol-LMC-SMC}
Generally, when comparing the Magellanic Cloud (MC) models to the solar
ones with identical input parameters one finds lower final outflow
velocities and lower dust-to-gas ratios. The radiative (in terms of the
gravitational) acceleration is weaker as well. For the mass-loss rate,
there are input parameters where the MC models show higher values than
the corresponding solar ones, as can be seen in
table~\ref{tab:ML-SMC-LMC-sol}.
\begin{table}
\caption{Mass-loss rates of SMC, LMC, and solar wind models with
  identical input parameters; $\epsilon_\text{C/O} = 1.8$, $\Delta v =
  5$~km s$^{-1}$, $P$ = 500 d at 8000 $L_\odot$, 600 d at 10000
  $L_\odot$, and 700 d at 12000 $L_\odot$. Mass loss averages
  are over time interval 70--150 periods, or (1) 70--90, (2) 70--210.}
\label{tab:ML-SMC-LMC-sol}
\centering
\begin{tabular}{ccr|lll}
$M$         & $T_\star$ & \multicolumn{1}{c|}{$L_\star$} &
  \multicolumn{3}{c}{$|\dot M|$[$10^{-5}M_\odot$yr$^{-1}$]} \\
$[M_\odot]$ & [K]      & \multicolumn{1}{c|}{[$L_\odot$]} & 
  SMC & LMC & solar \\ \hline
%M     T       L     smc    lmc    sol
0.8 & 2400 & 10000 & 2.19
                 $^{(2)}$ & 1.03 & 5.45 \\
0.8 & 2600 &  8000 & 2.75 & 3.31 
                        $^{(1)}$ & 2.38 \\
0.8 & 2600 & 10000 & 5.14 & 8.18
                        $^{(1)}$ & 3.57 \\
0.8 & 2800 & 10000 & 2.88 & 4.58 & 2.33 $^{(1)}$\\
0.8 & 2800 & 12000 & 4.69 & 6.74 & 3.23 \\
0.9 & 2800 & 10000 & 1.92 & 3.17
                        $^{(1)}$ & 1.94 \\
1.0 & 2600 & 10000 & 1.19 & 3.81
                        $^{(1)}$ & 2.19 $^{(1)}$ \\
\end{tabular}
\end{table}
The fact that one finds higher outflow velocities with increasing heavy
element abundance, and mass-loss rates of the same order of magnitude is
not surprising considering results from basic stellar wind theory. The
same behaviour is seen assuming an isothermal wind with an outward
directed $r^{-2}$ force applied outside the critical point. There the
mass-loss rate is set as long as the force is only acting in the
supersonic regime, that is outside the critical point.

Typically the outflow velocities of the solar metallicity models are
higher by about a factor of 2.2($\pm$0.2) than those of the LMC, and
4($\pm$1) than those of the SMC. The dust-to-gas ratios are higher by
about a factor of 1.3($\pm$0.1) and 2.3($\pm$0.2), respectively. For the
acceleration the trend is less clear, i.e.\ the data show a wider
spread, especially in the SMC case. The solar abundance models produce
values higher by a factor of 3.6($\pm$1.2), and 5.8($\pm$1.2),
respectively.

\begin{figure*}[tbp]
  \centering
  \includegraphics[width=0.95\textwidth]{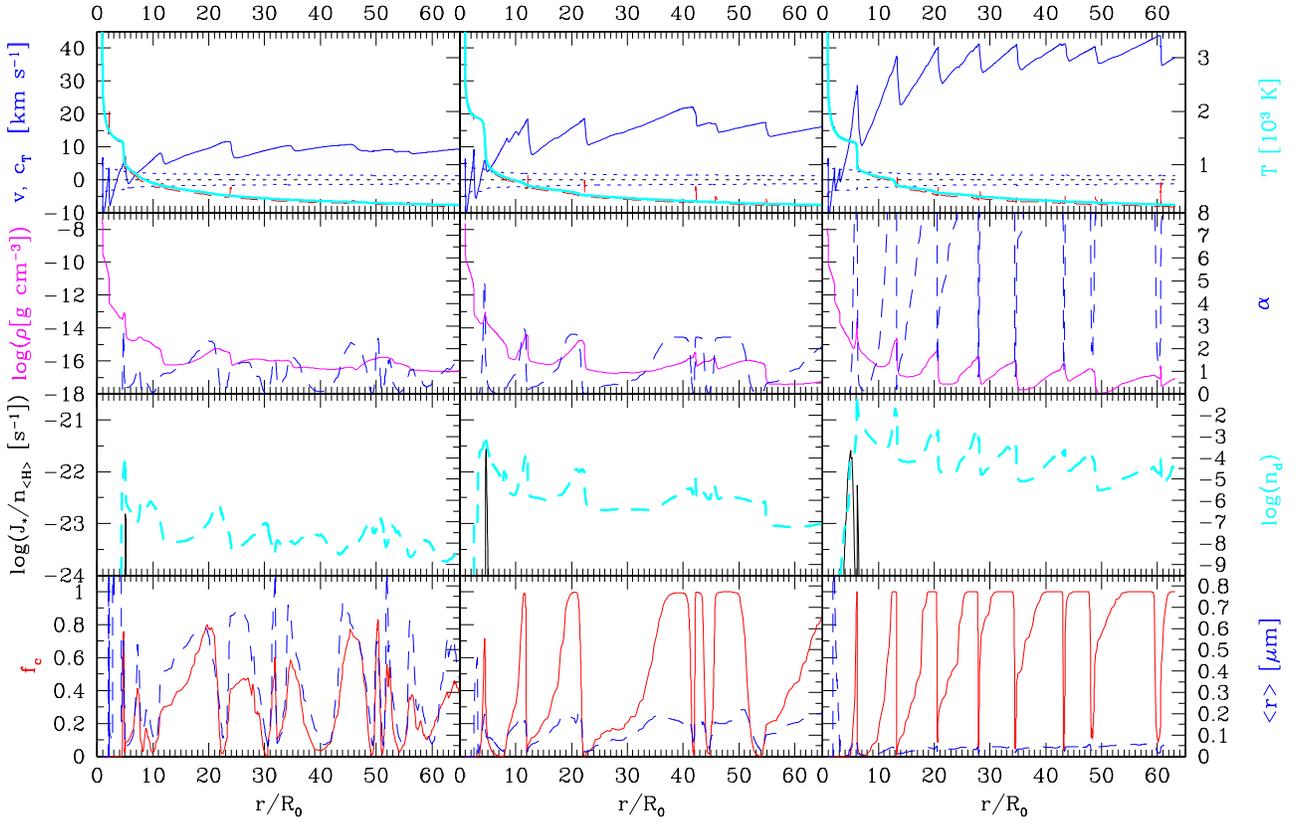}
  \caption{SMC (left), LMC (middle), and solar (right) model with identical
    input parameters $M = 0.9\,M_\odot$, $T_\star = 2800$~K, $L = 10^{4}
    L_\odot$, $P = 600$~d, $\Delta v = 5$~km~s$^{-1}$, $\epsilon_{\rm
    C}$/$\epsilon_{\rm O}=1.8$. Shown is the radial structure at time $t=90\,P$,
    $R_0$ being the stellar radius at the start of the calculation $t=0\,P$. The
    depicted quantities are (left scales) the radial velocity $v$ (solid blue
    line), the speed of sound $c_T$ (dotted blue lines -- in-/outwards, zero
    line as guide to the eye), the density $\rho$ (solid magenta), the
    nucleation rate per hydrogen atom $J_\star/n_{\langle H \rangle}$ (solid
    black), and the degree of condensation $f_\text{c}$ of the dust (solid red),
    (right scales) the gas temperature (dashed red) and equilibrium temperature
    (solid cyan) $T$, the ratio of radiative to gravitational acceleration
    $\alpha$ of the material (dashed blue), the dust density $n_\text{d}$
    (dashed cyan), and the mean dust particle radius $\langle r \rangle$ (dashed
    blue).}
  \label{fig:SMC-LMC-solar}
\end{figure*}

Figure \ref{fig:SMC-LMC-solar} depicts the radial structure of the model
with $M = 0.9\, M_\odot$, $T_\star = 2800$~K, $L = 10^{4} \, L_\odot$,
$P = 600$~d, $\Delta v = 5$~km~s$^{-1}$, and $\epsilon_{\rm
C}$/$\epsilon_{\rm O}$ = 1.8 for SMC, LMC, and solar abundances at the
instant of 90 periods after starting the model calculation. It gives a
good representation of these trends, especially for the outflow velocity
and the dust density, recalling that they are not only time but as well
radial averages. The acceleration trend is more indirectly represented,
since the graph is a time snapshot, but the acceleration average is a
time average of the value at the radius of the first maximum of the
condensation degree.

In the lowest panels two more trends can be seen. The condensation
degree which is closely related to the dust-to-gas ratio by mass
according to $\rho_\text{d}/\rho_\text{g} \propto \epsilon_\text{O}
(\frac{\epsilon_\text{C}}{\epsilon_\text{O}}-1)f_\text{c}$
\citep[see eq.~(4.7) of][]{winte1994} is less pronounced and more
irregular in LMC and SMC models than in solar models. While in this
solar metallicity model the condensation degree is virtually 1 in the
dust shells, the SMC model only reaches maximum value of about 0.8. That
is all the condensible material is accumulated in dust grains in the
solar model, but with lower metallicity the condensation is less
efficient. The second trend resulting from this simple graphical
comparison of respective models concerns the average grain size. In the
SMC case the grains reach much larger radii than with solar abundances.
Since the nucleation rate drops with decreasing metal abundance, there
are fewer grains in that case to collect the available C atoms.
Additionally, the grains have more time for growing due to their lower
velocity, hence they become larger.

\subsection{Approximative mass-loss formulae}\label{sec:formulae}

To give a mass-loss description applicable to stellar evolution we
restrict our model set to those of type A, describing stable
dust-driven wind. These are characterised by the criterion
$\langle\alpha\rangle_t > 1$.

Furthermore, one has to consider the treatment of the ``mechanical''
input parameters $\Delta v$ and $P$ as well as the C/O abundance ratio
for this purpose. Fortunately, the models are most sensitive to the
stellar parameters luminosity $L$, effective temperature $T$, and mass
$M$ \citep[for the solar metallicity models already pointed out
by][]{afs1997}.
Therefore, we followed for the LMC and SMC models the same approach as
for solar metallicity models \citep{wswas2002} and fixed the velocity
amplitude $\Delta v$ to a value of 5~km~s$^{-1}$. Setting the abundance
ratio $\epsilon_\text{C/O}$ to 1.8 allows for the fact that it has to be
high enough to produce a stable wind without being unreasonable.

Concerning the period $P$ we relied on observed period-luminosity
relations of LPVs. \citet{wswas2002} took the dependence of the
mass-loss rate on the period into account by transferring the slight $P$
dependence to an additional luminosity term. Now, we take the period
dependence more directly into account by choosing only those models
where the $P$-$L$ relation \citep{gw1996} is fulfilled. This is possible
since we now have a larger set of models at hand.

For the selected LMC models a multi-linear least square fit of the form $\log
\dot M = a_0 + a_1 * \log x_1 + \dots$ was performed leading to the following
formula:
\begin{eqnarray}\label{eq:lmcML}
|\dot M [M_\odot\,{\rm yr}^{-1}]| & = & 3.80 \times 10^{-5} \times (M_\star
[M_\odot])^{-2.56} \nonumber \\ & & \times \left(\frac{T_\star [{\rm
      K}]}{2600}\right)^{-7.44} \times \left(\frac{L_\star
    [L_\odot]}{10^4}\right)^{2.86}
\end{eqnarray}
with a correlation coefficient of $K = 0.98$.

For the SMC models the result is:
\begin{eqnarray}\label{eq:smcML}
|\dot M [M_\odot\,{\rm yr}^{-1}]| & = & 2.34 \times 10^{-5} \times (M_\star
[M_\odot])^{-3.01} \nonumber \\ & & \times \left(\frac{T_\star [{\rm
      K}]}{2600}\right)^{-6.22} \times \left(\frac{L_\star
    [L_\odot]}{10^4}\right)^{2.84}
\end{eqnarray}
with a correlation coefficient of $K = 0.97$.

The formula for solar element composition as published in \citet{wswas2002}
reads, for comparison:
\begin{eqnarray}\label{eq:solML}
|\dot M [M_\odot\,{\rm yr}^{-1}]| & = & 4.52 \times 10^{-5} \times (M_\star
[M_\odot])^{-1.95} \nonumber \\ & & \times \left(\frac{T_\star [{\rm
      K}]}{2600}\right)^{-6.81} \times \left(\frac{L_\star
    [L_\odot]}{10^4}\right)^{2.47}
\end{eqnarray}
where the correlation coefficient is $K = 0.97$.

As already pointed out for the solar metallicity case, the most
influential parameter in the formulae for LMC/SMC abundances is again
the effective temperature. This reflects the extreme sensitivity of the
dust nucleation with respect to the local temperature.

%%% Abb: fitMLR over T, L. M fixed %%%%%
\begin{figure}[tbp]
  {\noindent\small {\bf (a)} LMC:}\\[-2ex]
    \includegraphics[angle=270,width=0.49\textwidth]{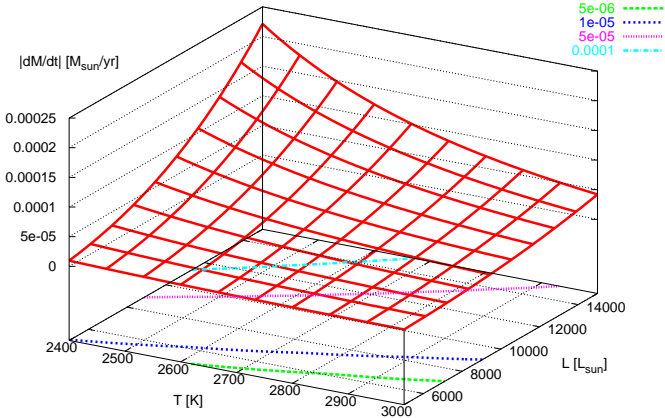}
\\[2ex]
  {\noindent\small {\bf (b)} SMC:}\\[-2ex]
  \includegraphics[angle=270,width=0.49\textwidth]{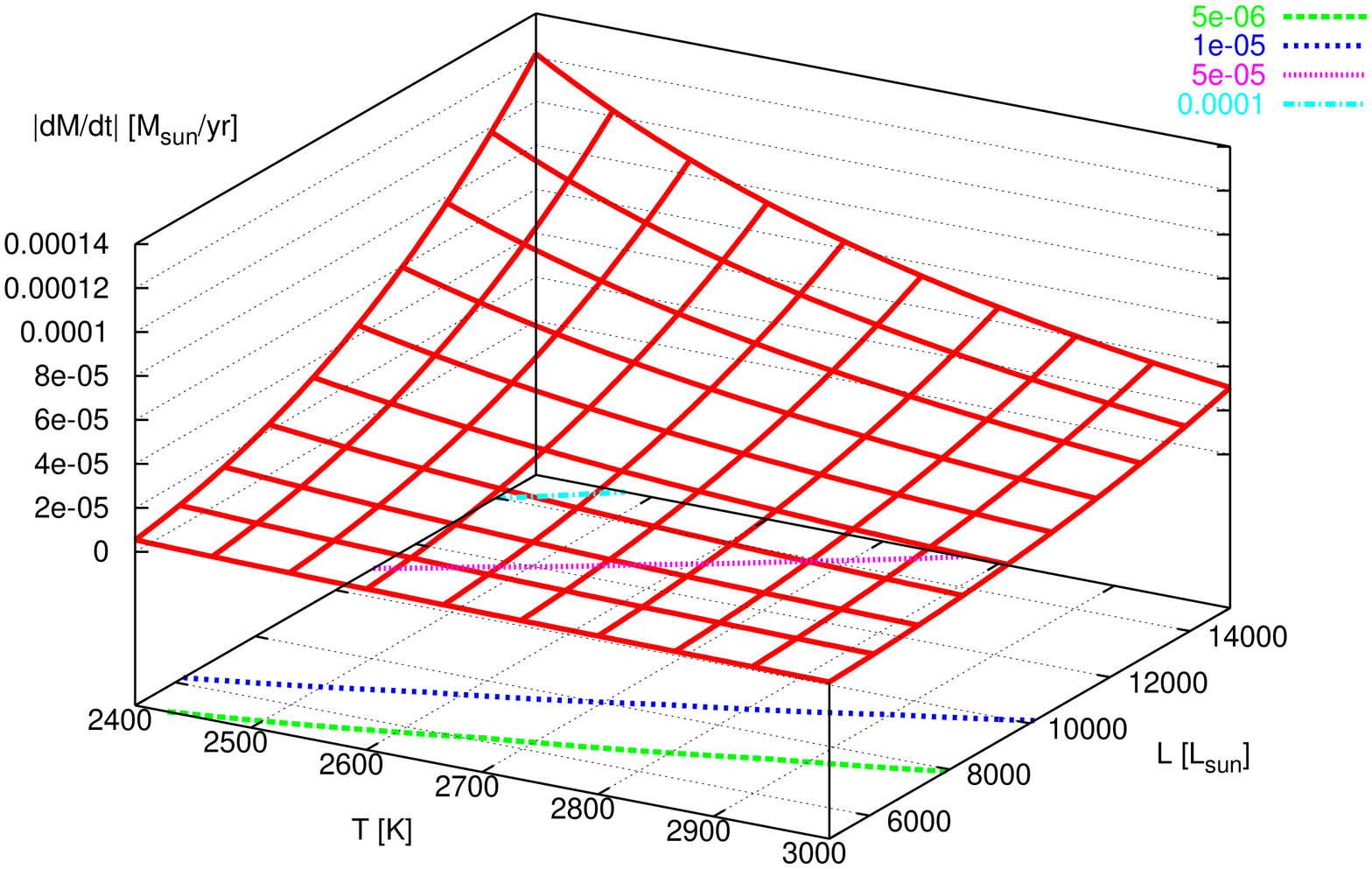}
  \caption{Graphical illustration of the fit formulae for the mass loss rate for
    fixed mass $M_\star = 1.0\, M_\odot$.}
  \label{fig:massLoss}
\end{figure}
Graphical illustrations of the formulae for LMC and SMC are given in
figure~\ref{fig:massLoss} for a fixed mass of $M_\star = 1.0\, M_\odot$,
temperature range of 2400 $\dots$ 3000~K, and luminosity of 5000 $\dots$
15000~$L_\odot$.  The lines on the base of the diagrams are projections
of contour lines, i.e.\ they mark those temperature and luminosity
values for which the resulting mass-loss rate is $10^{-4}$, $5 \times
10^{-5}$, $10^{-5}$, and $5 \times 10^{-6}$~$M_\odot$~yr$^{-1}$,
respectively. These show that for the SMC higher luminosities / lower
temperatures are necessary to reach the same mass-loss rate as in the
LMC.

To give the validity limit of the above formulae the selection criterion
needs to be expressed in terms of the stellar parameters, since $\langle
\alpha \rangle$ is a quantity which does not appear in stellar evolution
calculations. By considering the ratio $L/M$ over temperature $T$ for
each metallicity set it turns out that the criterion can be approximated
using $\frac{L}{M} \propto T^x$.  We find that
equations~(\ref{eq:lmcML}) and (\ref{eq:smcML}) are valid for
\begin{equation}
  \label{eq:critLum}
  L > L_\text{crit} \quad , \quad L_\text{crit} = \left\{ \begin{array}{ll}
      5.58\times10^{-8} \, T^{3.198} \, M , &  \text{LMC} \\
      5.47\times10^{-4} \, T^{2.126} \, M , &  \text{SMC}
    \end{array} \right.
\end{equation}
The need to include such precise values of the temperature exponent expresses 
again the steep dependence of the dust nucleation rate on the local kinetic 
gas temperature.

%__________________________________________________________________

\section{Results from application to stellar evolution}

In order to get some insight into the overall mass loss of subsolar
metallicity stars on the AGB compared their solar counterparts, we
applied the derived mass-loss rates to stellar evolution calculations.
For that purpose, we used again the well-calibrated Cambridge evolution
code \citep[cf.][and references therein]{wswas2002}, as we did for the
solar metallicity models. A particular advantage of these evolution
models is a good match of their effective temperatures for stars high on
the AGB, according to a semi-empirical adjustment made by
\citet{sws1999}.
This is an important point: Only a small mismatch in $T_{\rm eff}$ can
result in a significantly different mass-loss rate, despite the use of
the very same mass-loss formula, because of the high dependence of the
mass-loss on $T_{\rm eff}$ of the stellar models.

Mass loss is included by reducing the stellar mass at the outer
boundary, whereby the particular formulation depends on the current
evolutionary stage of the model. Mass loss is considered only after the 
star runs out of hydrogen in the core and reaches the red giant branch
(RGB). The formulae derived in section~\ref{sec:formulae} are valid
above the respective critical luminosities. These are reached when the
star has developed up to the tip of the AGB and has entered the
thermally pulsing phase. Before this phase we apply a formula given by
\citet{sc2005} who have reconsidered the well-known Reimers formula
\citep{reime1975} by using theoretically motivated arguments. As a
result it now includes two new terms containing a dependence on the
effective temperature and the surface gravity, respectively. Around
$L_\text{crit}$ a short transition zone is introduced where the code
interpolates between the modified Reimers and the formula based on
dust-driven wind models to avoid artificial jumps in the mass-loss rate.

We have generated model grids with a metallicity of $Z = 0.01$ and $Z =
0.001$ to represent LMC and SMC stars, respectively, the closest values
for which appropriate necessary input data is available.

Table~\ref{tab:totML} lists the initial masses and the total mass lost
over the whole life of the star, according to a number of our evolution
models with mass loss for subsolar metallicity. Within each grid we
calculated evolutionary tracks without and with overshooting \citep[with
a value of $\delta_\text{ov}=0.12$ for the overshoot parameter as tested
by][]{spe1997}.

\begin{table*}[tbp]
  \caption{Stellar evolution grids: final masses and ages}
  \label{tab:totML}
  \centering
  \begin{tabular}{l|rr|rr||rr|rr}
    & \multicolumn{4}{c||}{Z=0.01} & \multicolumn{4}{c}{Z=0.001} \\
                 & \multicolumn{2}{c|}{$\delta_\text{ov}=0$}&
                 \multicolumn{2}{c||}{$\delta_\text{ov}=0.12$} &
                 \multicolumn{2}{c|}{$\delta_\text{ov}=0$} &
                 \multicolumn{2}{c}{$\delta_\text{ov}=0.12$} \\ \hline
    $M_\text{i}$ & $M_\text{e}$ & age & $M_\text{e}$ & age & $M_\text{e}$ & age
    & $M_\text{e}$ & age \\
    $[M_\odot]$ & $[M_\odot]$ & [10$^{9}$yr] & $[M_\odot]$ & [10$^{9}$yr] &
    $[M_\odot]$ & [10$^{9}$yr] & $[M_\odot]$ & [10$^{9}$yr] \\ \hline \hline
    0.70 &      &       &      &       & 0.47 & 24.24 & 0.47 & 24.33  \\
    0.75 &      &       &      &       & 0.49 & 18.83 & 0.49 & 18.91  \\
    0.80 & 0.47 & 23.77 & 0.47 & 23.79 & 0.50 & 14.88 & 0.50 & 14.95  \\
    0.85 & 0.48 & 19.02 & 0.48 & 19.04 & 0.53 & 12.06 & 0.54 & 12.14  \\
    0.90 & 0.49 & 15.41 & 0.49 & 15.43 & 0.55 &  9.85 & 0.56 &  9.92  \\
    0.95 & 0.53 & 12.73 & 0.53 & 12.73 & 0.57 &  8.15 & 0.57 &  8.21  \\ \hline
    1.00 & 0.55 & 10.53 & 0.55 & 10.54 & 0.58 &  6.82 & 0.59 &  6.88  \\
    1.05 & 0.56 &  8.80 & 0.56 &  8.83 & 0.59 &  5.78 & 0.60 &  5.83  \\
    1.10 & 0.57 &  7.42 & 0.57 &  7.48 & 0.59 &  4.94 & 0.60 &  4.99  \\
    1.15 & 0.58 &  6.31 & 0.58 &  6.40 & 0.60 &  4.26 & 0.61 &  4.32  \\
    1.20 & 0.59 &  5.42 & 0.58 &  5.53 & 0.61 &  3.70 & 0.62 &  3.77  \\
    1.25 & 0.59 &  4.70 & 0.59 &  4.80 & 0.62 &  3.24 & 0.63 &  3.31  \\
    1.30 & 0.60 &  4.11 & 0.59 &  4.19 & 0.63 &  2.86 & 0.64 &  2.93  \\
    1.35 & 0.60 &  3.61 & 0.60 &  3.69 & 0.63 &  2.54 & 0.64 &  2.61  \\
    1.40 & 0.61 &  3.19 & 0.60 &  3.26 & 0.64 &  2.26 & 0.65 &  2.34  \\
    1.45 & 0.61 &  2.85 & 0.61 &  2.92 & 0.65 &  2.03 & 0.66 &  2.11  \\ \hline
    1.50 & 0.62 &  2.55 & 0.61 &  2.62 & 0.66 &  1.83 & 0.67 &  1.92  \\
    1.55 & 0.62 &  2.29 & 0.61 &  2.37 & 0.66 &  1.65 & 0.67 &  1.75  \\
    1.60 & 0.63 &  2.07 & 0.62 &  2.16 & 0.67 &  1.50 & 0.68 &  1.61  \\
    1.65 & 0.63 &  1.87 & 0.63 &  1.98 &      &       & 0.69 &  1.49  \\
    1.70 & 0.64 &  1.70 & 0.63 &  1.82 &      &       & 0.69 &  1.38  \\
    1.75 & 0.64 &  1.55 & 0.64 &  1.67 &      &       & 0.70 &  1.35  \\
    1.80 & 0.65 &  1.43 & 0.64 &  1.55 &      &       & 0.70 &  1.27  \\
    1.85 & 0.65 &  1.32 & 0.65 &  1.44 &      &       & 0.71 &  1.19  \\
    1.90 & 0.66 &  1.22 & 0.66 &  1.37 &      &       & 0.72 &  1.10  \\
    1.95 & 0.66 &  1.14 & 0.66 &  1.46 &      &       & 0.73 &  1.03  \\ \hline
    2.00 & 0.67 &  1.06 & 0.67 &  1.36 &      &       & 0.74 &  0.96  \\
    2.05 & 0.67 &  1.00 & 0.67 &  1.27 &      &       & 0.75 &  0.90  \\
    2.15 & 0.68 &  0.90 & 0.68 &  1.12 &      &       & 0.77 &  0.79  \\
    2.25 & 0.69 &  0.82 & 0.69 &  0.99 &      &       & 0.79 &  0.70  \\
    2.35 &      &       & 0.71 &  0.88 &      &       & 0.81 &  0.62  \\
    2.50 &      &       &      &       &      &       & 0.85 &  0.53 \\
  \end{tabular}
\end{table*}

\subsection{Mass loss histories}\label{sec:MLhist}

\begin{figure*}[tbp]
  {\noindent\small {\bf (a)} $Z = 0.01$, $M_\text{i} = 1.25 \, M_\odot$:}\hfill
  \makebox[0.5\textwidth][l]{\small {\bf (b)} $Z = 0.001$, $M_\text{i} = 1.95 \, M_\odot$:}\\
  \includegraphics[angle=270,width=0.49\textwidth]{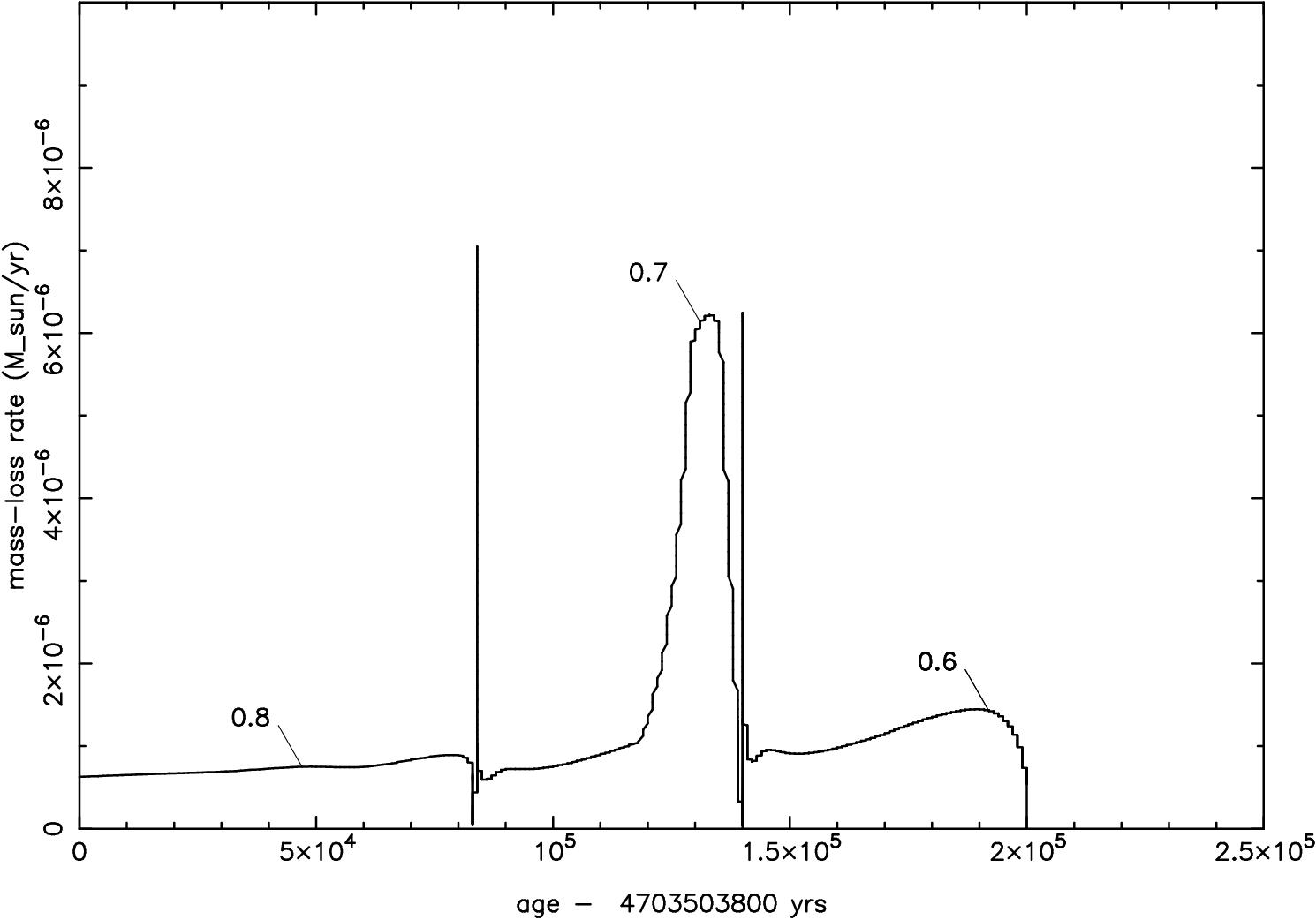}
  \hfill
  \includegraphics[angle=270,width=0.49\textwidth]{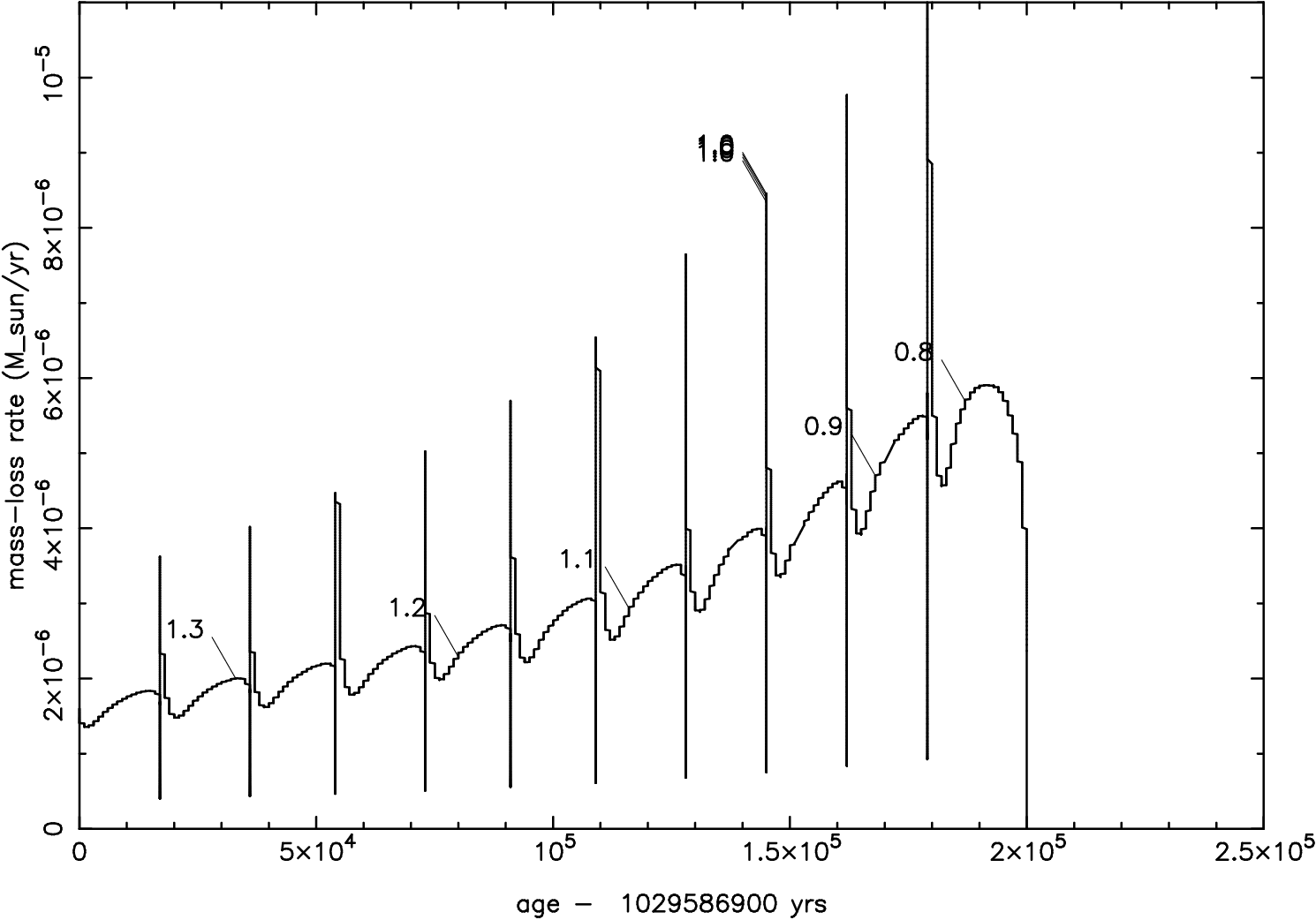}
  \\[1ex]
  {\small {\bf (c)} $Z = 0.01$, $M_\text{i} = 2.50 \, M_\odot$:}\hfill
  \makebox[0.5\textwidth][l]{\small {\bf (d)} $Z = 0.001$, $M_\text{i} = 2.50 \, M_\odot$:}\\[-1ex]
  \includegraphics[angle=270,width=0.49\textwidth]{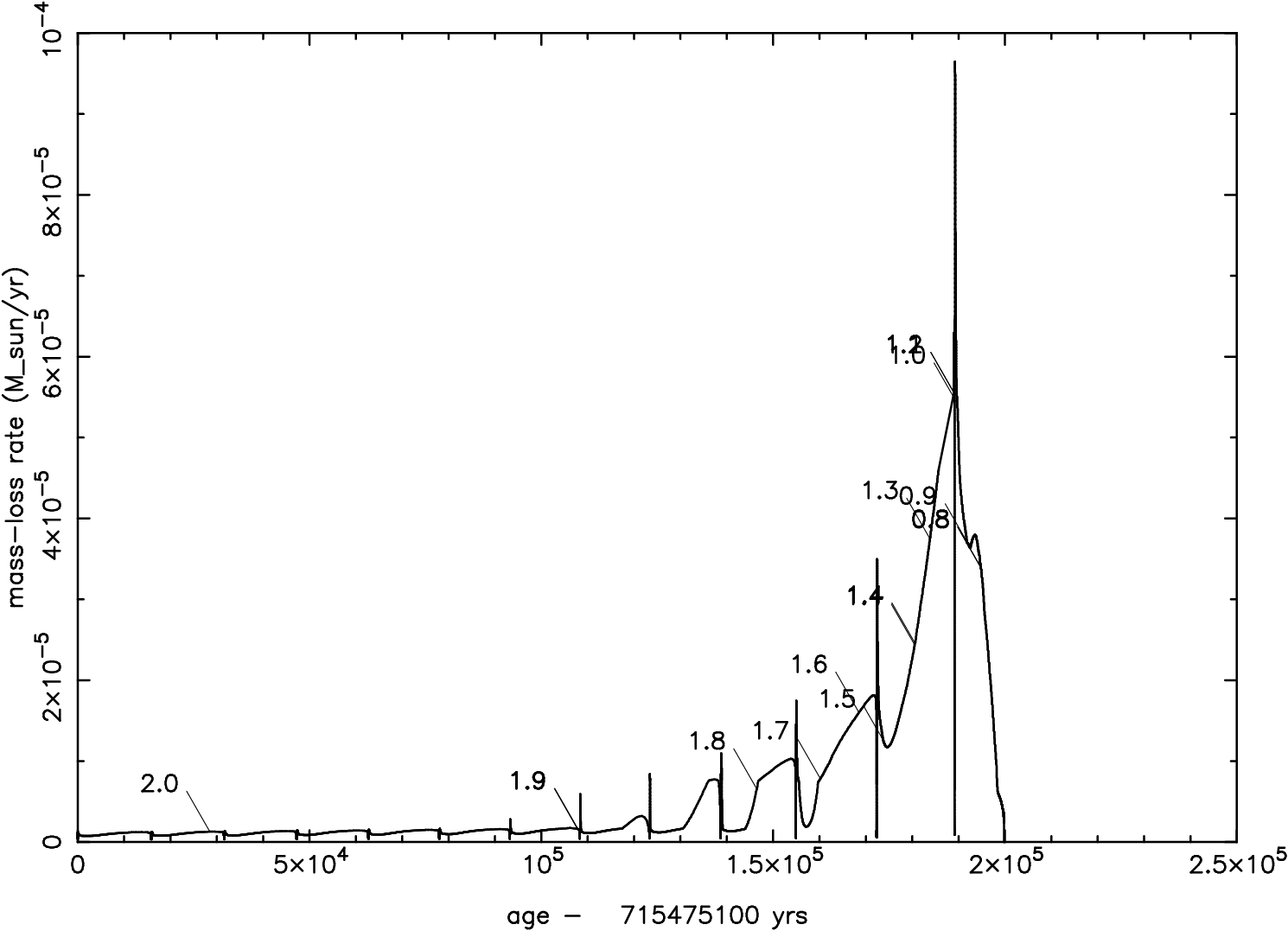}
  \hfill
  \includegraphics[angle=270,width=0.49\textwidth]{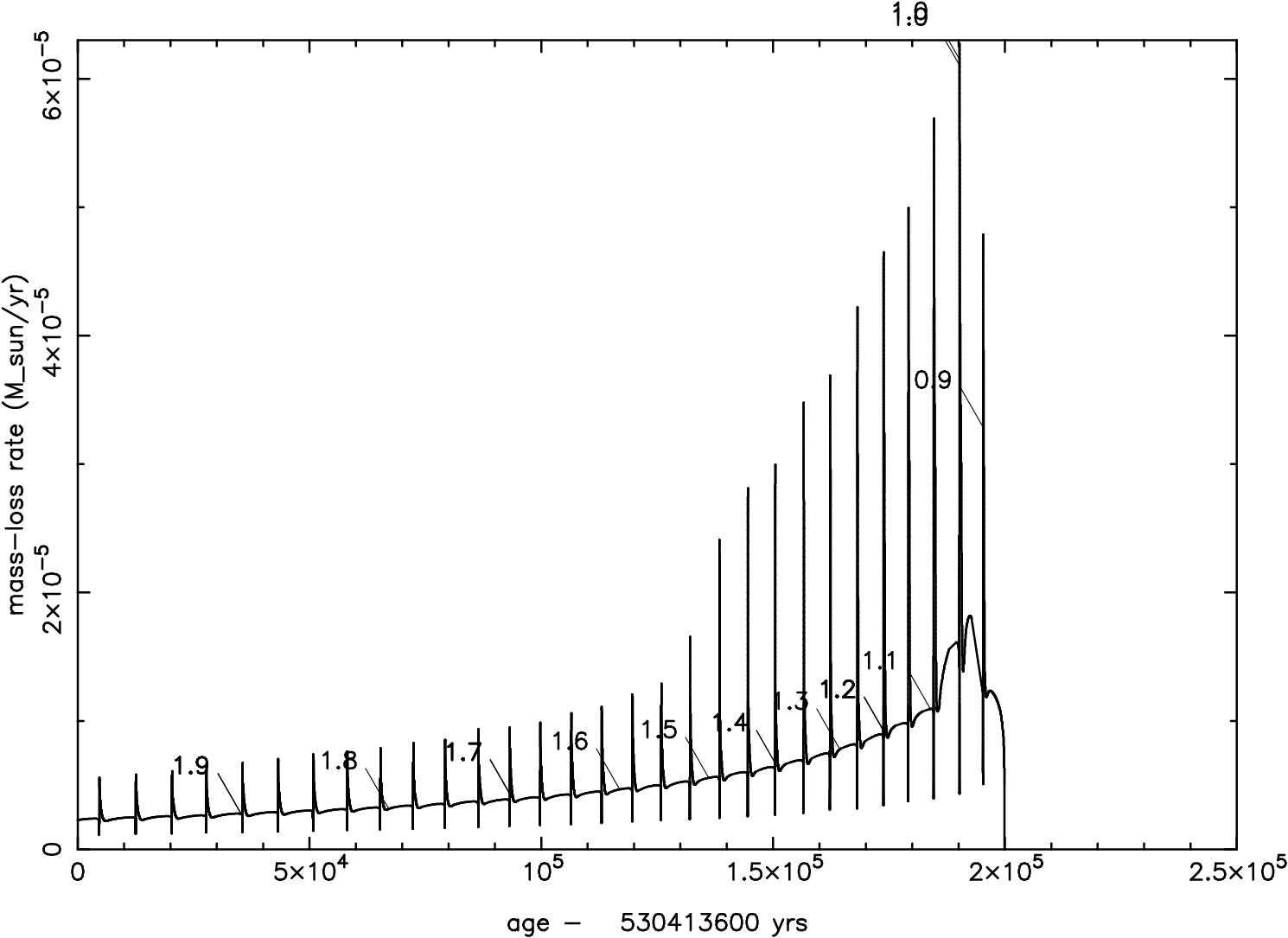}
  \caption{Mass loss histories of stars with different initial masses and
    metallicities. Labels denote the current mass.}
  \label{fig:MLhistLMC}
\end{figure*}

The temporal development of the stellar mass loss rate is of particular
interest, since the density structure of a planetary nebula is
determined by the mass-loss history of its progenitor star.
Figure~\ref{fig:MLhistLMC} shows four examples of such mass-loss
histories.
Depicted are the rates of $Z=0.01$ models with an initial mass of (a)
$M_\text{i} = 1.25 \, M_\odot$ and (c) $M_\text{i} = 2.50 \, M_\odot$,
and $Z = 0.001 \, M_\odot$ models with (b) $M_\text{i} = 1.95 \,
M_\odot$ and (d) $M_\text{i} = 2.50 \, M_\odot$. Each graph is plotted
over the final 200\,000 years of the stellar life, but the scales of the
mass-loss rates differ! The graphs (a) and (b) show the results for the
lowest initial mass where the critical luminosity is reached and a brief
onset of a dust-driven wind results, respectively, while graphs (c) and
(d) show the highest initial mass where the evolution is followed until
the star leaves the AGB.

What catches the eye at first glance is the fact that the number of
thermal pulses (seen as spikes)
increases and the inter-pulse duration decreases with
decreasing metallicity (cf.\ figures~\ref{fig:MLhistLMC}(c) and
(d)). This finding is in accordance with the predictions of other
stellar evolution calculations using different mass loss prescriptions,
see for example \citet{marig2001}, and references therein. As pointed
out there, the exact number of thermal pulses depends on the chosen
mass-loss formalism.

Another clear difference is the overall shape of the mass-loss histories
of models with different metallicity: The $Z = 0.01$ graph resembles the
mass-loss rates of our models with solar metallicity, see figures~2--5
in \citet{wswas2002}. For these metallicities the critical luminosities
are fairly similar and the stars reside in a regime above that value
while they are in the thermally pulsing AGB phase. As a result the mass
loss in this late evolutionary stage is controlled by our formula based
on dust-driven winds. By contrast, the $Z = 0.001, M = 2.5\,M_\odot$
rate is mostly dominated by not dust-driven mass-loss, here described by
the modified Reimers law. Only during the last few thermal pulses does
the star reach luminosities high enough for the dust-driven wind formula
to apply. Our present models for SMC-like metallicity therefore result
in a rather smooth and broad density distribution of the blown-out
material developing to a planetary nebula\footnote{Naturally, this is a
simplified picture, since it disregards different velocities with which
the material might be driven away.}. Models with solar and LMC
abundances though have distinctly higher mass-loss rates during the last
$\sim$25\,000 years leading to a denser and narrower mass distribution.

\subsection{Total masses lost}\label{sec:intML}

The total mass lost by a star in the course of its evolution, for given
initial stellar mass, is an important quantity in modelling the galactic
chemical evolution. A further quantity of interest is the mass of the
resulting white dwarf (WD) or, rather, the WD mass distribution, since
WDs are considered to be the end-products of stellar evolution for
intermediate masses. The final stellar (WD) mass is basically identical
with the core mass of the AGB star supposing that the whole envelope is
ejected via stellar winds. Therefore we present in the following the
total (integrated over the whole stellar life) masses lost by stars with
subsolar metallicity, according to our models.

\begin{figure}[tbp]
  {\small {\bf (a)} Models with metallicity $Z=0.01$:}\\[-1ex]
  \centerline{
    \includegraphics[angle=270,width=0.49\textwidth]{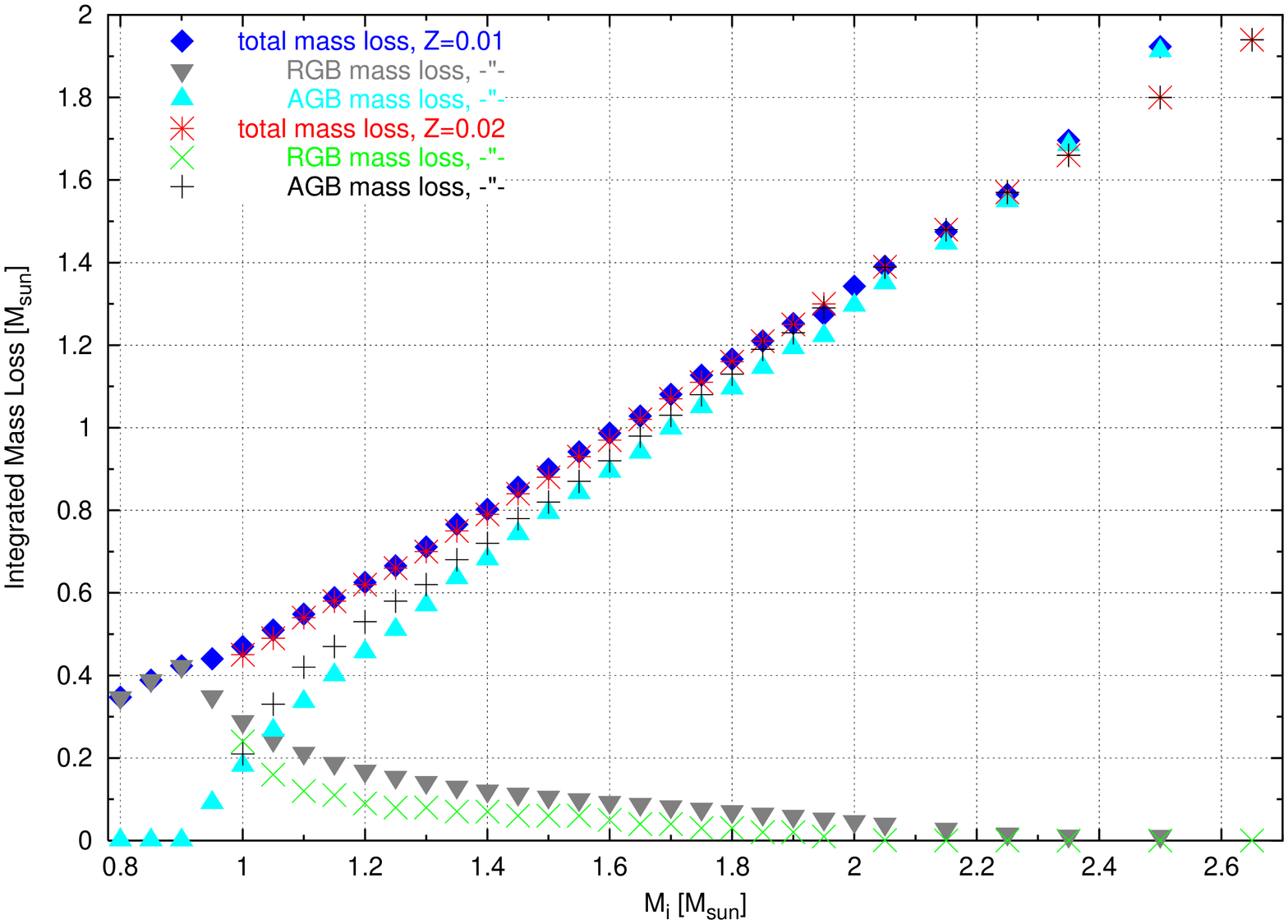}
  }
  \\
 {\small {\bf (b)} ...and $Z=0.001$:}\\[-1ex]
  \centerline{
    \includegraphics[angle=270,width=0.49\textwidth]{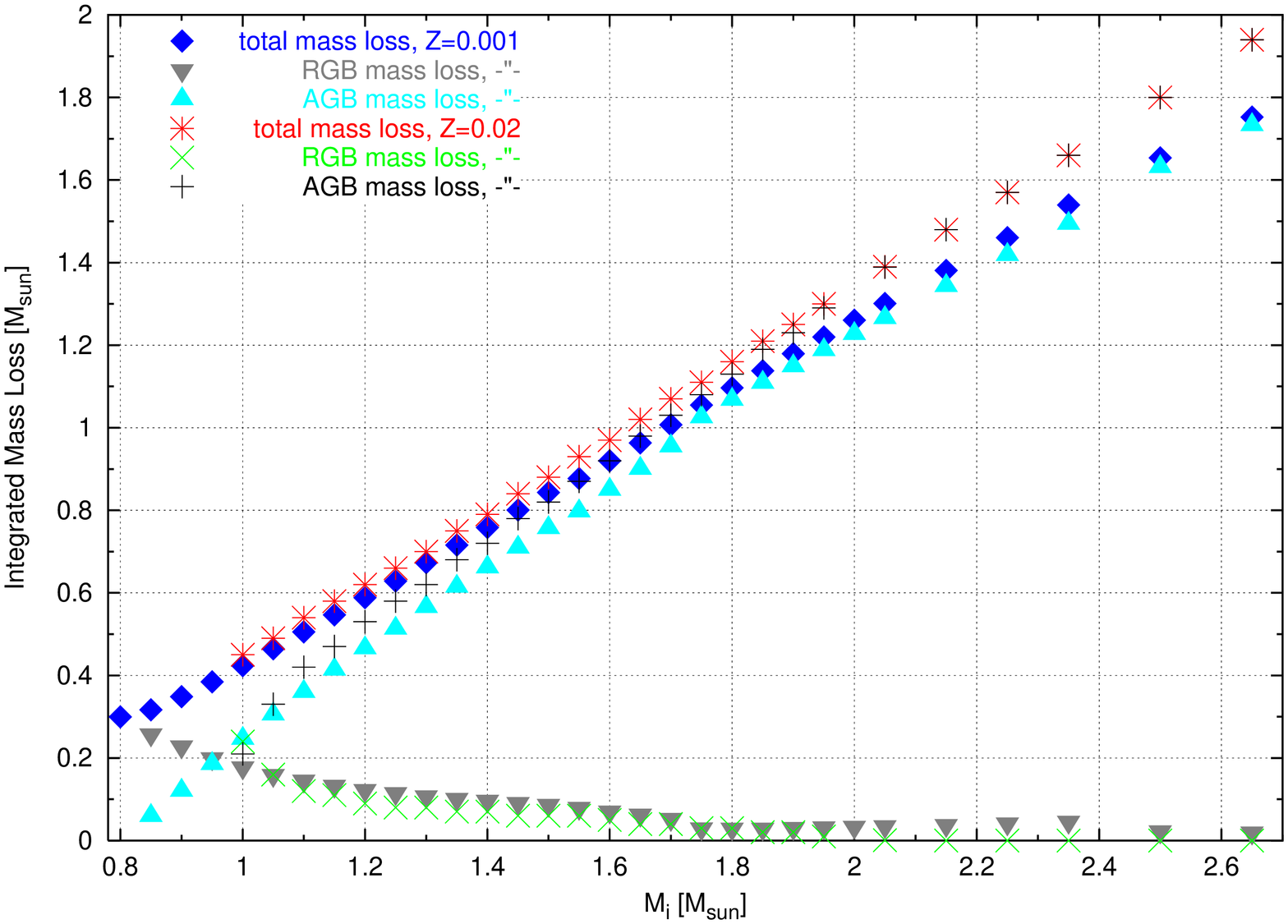}
  }
  \caption{Integrated mass loss as a function of
    initial mass. For each initial mass the total mass loss is shown
    (blue diamonds) as well as the respective fractions lost on the
    RGB (cyan) and the AGB (grey). For reference the respective solar
    values are given as well in both plots (red, green, and black
    crosses).}
  \label{fig:totML}
\end{figure}

Figure~\ref{fig:totML} depicts the amount of mass lost by a star of
given initial mass $M_\text{i}$ during its life for the two grids of
tracks with metallicity (a) $Z=0.01$ and (b) $Z=0.001$. In addition, we
detail the time-integrated mass-loss of the RGB and AGB phases. In both
graphs the respective masses lost by models with solar metallicity (and
same initial mass) are plotted as well, for comparison.
It can be seen in figure~\ref{fig:totML}(a) that the total amount of
mass loss is virtually the same, but not lower for LMC-like than for
solar abundances. However, the contributions lost on the RGB and AGB are
shifted, especially for the lower initial masses shown more mass is
already lost in the RGB phase. For $Z=0.001$ figure~\ref{fig:totML}(b)
shows that the total amount of mass lost is almost the same as for solar
metallicity for the lower end of initial masses. With higher initial
mass though, less material is blown away for SMC-like abundances. Again
there is a shift towards higher contribution of the RGB, but less
pronounced than in case~(a).

%__________________________________________________________________

\section{Discussion}

In order to account for carbon-rich chemistry, we enhance the carbon element
abundance considering the C/O ratio as a model parameter. In this work we
considered mainly values of 1.3 and 1.8, the latter especially in models with
SMC abundances. This value might be regarded as rather extreme. Nevertheless,
\citet{mzlym2005} reported possible C/O values of $\sim$1.4 for LMC
stars and even higher for the SMC, resulting from a chemical equilibrium
model to explain the strength of observed C$_2$H$_2$ absorption
features.

\Citet{vczl2005}
observed AGB stars in the Large Magellanic Cloud, C-stars as well as
M-type stars, and derived mass-loss rates of $\log \dot
M$[$M_\odot$~yr$^{-1}$] $\sim$ -5.6 $\ldots$ -5 for C-type and $\sim$ -5
$\ldots$ -4.2 for M-type giants. Furthermore, they give an empirical
mass-loss formula for these stars dependent on effective temperature and
luminosity and compare their formula to equation~(\ref{eq:solML})
derived from our models with solar abundances. Even though the
dependence of our LMC formula on effective temperature is stronger, it
is consistent with their observed value within the errors. Furthermore,
they argued that even if their observed luminosity dependence is much
lower than our theoretical value, they may still be consistent as the
effect of decreasing mass-loss rate with increasing stellar mass
counteracts the luminosity dependence.

An issue which remains uncovered by the models presented here is
related to the assumption of grey radiative transfer and its effect on
the resulting mass-loss rates. \citet{hgaj2003} included
frequency-dependent treatment in their models and found those to have
cooler and denser upper layers which is in favour of a more efficient
dust condensation. A radiative pressure 2--3 times larger than the grey
values results, and the mass-loss rates and wind velocities are larger
in the non-grey case for the three models compared to grey ones
calculated with Planck-mean opacities. On the other hand,
\citet{hws2000}
investigated the effects of molecular opacities on the characteristics
of carbon-rich dust driven wind models in the frame work of grey
opacities. These authors compared opacities represented by the same
constant value used here, with Rosseland-mean and Planck-mean
averages. As a general result it is found that the mass loss rates and
wind velocities are highest for the constant value of the opacity also
used here and lowest for the Planck-mean opacities, whereas the
Rosseland-mean leads to intermediate values of the mass loss rate.
Considering these results and the results of \citet{hgaj2003} it is to
be expected that the inclusion of non-grey opacities in our models may
lead to somewhat reduced mass-loss rates and wind velocities. A
quantitative assessment of the effect would however require the
calculation of an appropriate set of models with non-grey treatment of
the gas opacities for the sub-solar element abundances discussed here.

Concerning the comparison of hydrodynamical wind models with different
element compositions \citet{mwhe2008}
suggest that one should rather keep the abundance difference C-O fixed
than the ratio C/O, since for dust formation the amount of free carbon
is important assuming complete CO blocking. Doing so effectively leads
to higher C/O ratios for lower metallicity where the oxygen abundance is
decreased as well. For the mass-loss rates this different approach of
comparing models consequently has not much influence, since the models
with higher C/O are even further beyond the threshold. For example, we
calculated the LMC and SMC model shown in figure~\ref{fig:SMC-LMC-solar}
with fixed C-O
($\log_{10}(\epsilon_\text{C}-\epsilon_\text{O})\sim$8.72)\footnote{which
corresponds to values of C/O~$\sim$ 3.4 and 5.9, respectively}
instead C/O. This leads to averaged mass-loss rates reduced by a factor
of 0.7 and raised by a factor 1.3 (to the value with fixed C/O),
respectively. Considering the hydrodynamical structure itself, the
models with fixed C-O are very similar to their solar counterparts,
e.g.\ they show velocities of the same order and the degree of
condensation reaches 100\% in the dust shells even in the SMC model.

Despite the differences in the assumptions of gas opacities and
radiative transfer we come to the same conclusion about mass loss as
\citet{mwhe2008}: Lower metallicities do not necessarily result in lower
mass-loss rates.

\mbox{}

As stated before the wind models predict higher critical luminosities
for a dust-driven wind to develop for SMC abundances than for the LMC,
e.g.\ a factor of 1.7 at $\log T = 3.5$. On the other hand, stellar
evolution tracks of stars with $Z = 0.001$ reach higher luminosities (up
to 0.2 dex) at the tip of the AGB than their $Z = 0.01$ counterparts
with same initial mass. In other words our models predict SMC stars to
undergo mass loss due to dust-driven winds, at least those with initial
masses of about two solar masses and higher, despite the higher critical
luminosity required to do so.

%__________________________________________________________________

\section{Conclusions}

In this article we present dust-driven wind models with subsolar
metallicities as observed in the Magellanic Clouds. These are calculated
using an adapted version of the hydrodynamical code developed at the
Technical University Berlin, with a radiative transport description
improved for lower opacities. Based on a grid of models for SMC and LMC
metallicity respectively, we derived approximative mass-loss formulae
valid above a critical luminosity, that is on the tip of the AGB. In
order to investigate the overall stellar mass loss we applied these
descriptions to stellar evolution using the Cambridge code.

\mbox{}

Our main results can be summarised as follows:
\begin{enumerate}
\item A comparison of the subsolar metallicity models to the
corresponding solar ones with fixed C/O ratio reveals that typically the
outflow velocities of the solar models are higher by about a factor of
2.2 than those of the LMC, and a factor of 4 higher than those of the
SMC. This is because in the solar metallicity case a larger amount of
dust is formed which can more efficiently accelerate the wind outside of
the sonic region.
\item The averaged mass-loss rates of these models are of the same
order of magnitude as the solar ones. This can be understood by the
threshold character of the radiative acceleration in a dust-driven wind:
Only a moderate amount of dust is needed to form in the sub-sonic wind
region to accelerate the wind to super-sonic velocities (equivalent to
$\alpha > 1$ at sub-sonic velocities). The mass-loss rate is in this
case determined in the sub-sonic region, irrespective of how much
additional dust is still formed further out in the wind.
\item The most influential parameter in the mass-loss formulae is the
temperature hinting at the sensitive dependence of the dust formation to
this quantity.
\item Once the abundance ratio C/O is above a lower limit it still has
an impact on the hydrodynamical structure of the models, e.g.\ different
outflow velocities, but the averaged mass-loss rate does not change
dramatically anymore. Again this is plausible due to characteristics of
the driving force in a radiation driven wind (see also
section~\ref{sec:sol-LMC-SMC}).
\item While LMC and solar abundance wind models result in fairly equal
critical luminosities necessary to drive a wind by radiation pressure on
dust, higher values are required in the SMC models. Since stellar
evolution models with lower metallicity reach higher luminosities on the
tip of the AGB, some of these stars (with initial masses $M_\text{i}
\gtrsim 2 M_\odot$) suffer dust-driven mass loss nevertheless.
\end{enumerate}

%__________________________________________________________________

\begin{acknowledgements}
  We thank S.~H{\"o}fner and L.~Mattsson for discussions about their
  hydrodynamical models in general, and for insights into effects
  of non-grey radiative transfer in particular.
  Part of this work was supported by the German \emph{Deut\-sche
    For\-schungs\-ge\-mein\-schaft, DFG\/} project number Se 420/22.
  KPS would like to acknowledge support by Conacyt (grant No. 57744).
\end{acknowledgements}

\end{document}